\documentstyle[preprint,aps]{revtex}

\newcommand{\lsim}[1]{
\setlength{\unitlength}{12pt}
\begin{picture}(1.4,1.)
\put(.7,-0.3){\makebox(0.0,1.)[t]{$<$}}
\put(.7,-0.3){\makebox(0.0,1.)[b]{$\sim$}}
\end{picture}#1}
\begin{document}
\draft

\title{Observable interactions of quintessence with ordinary matter and
neutrinos}

\author{R. Horvat \\
   ``Rudjer Bo\v skovi\' c'' Institute, P.O.Box 180, 10002 Zagreb,
Croatia}

\maketitle

\begin{abstract}

For any realistic theory of quintessence that allows for a violation of the
equivalence principle (VEP), we study the experimental constraints on such
theories coming from ordinary matter as well as neutrinos. We discuss and 
compare constraints from two basically different (extra) contributions to
differences of fermion masses: one is due to the vacuum expectation value
(vev) of quintessence, and the other is based on a quintessence-exchange
graph at finite temperature/density.
\end{abstract}

\newpage

There are now increasing indications,
based on recent redshift-distance measurements of
High-Z Supernovae Ia \cite{1},measurements of different observables on rich
clusters of galaxies which all point towards a low value of the fraction
density in matter \cite{2} as well as recent Cosmic Microwave Background
measurements \cite{3}, that the Universe is presently undergoing
accelerated expansion due to a smooth component with negative pressure,
sometimes also called `dark energy' \cite{4}. For a long time the simplest
candidate has been a nonvanishing cosmological constant. Such a  situation has
changed recently, when a dynamical, slowly-rolling, spatially inhomogeneous
scalar field component, named `quintessence' \cite{5}, had been put forward
as an alternative candidate. Generally, models of quintessence are currently
better motivated; because of more parameters involved in such models they
have proven more useful in explaining some of the difficulties left by the
cosmological constant scenarios. Besides the `cosmological constant problem'
\cite{6}, the quintessence models face the `why now' problem:
since the dark energy density and the matter energy density decrease at
different rates as the Universe expands, it is not clear why we appear to
live in an era during which the two energy densities are roughly the same,
with the concordant values, $\Omega_{\Lambda} \sim 2/3 $, $\Omega_{m} \sim
1/3 $ \cite{7}, $\Omega_{m}$ and $\Omega_{\Lambda}$ being the fraction
densities in matter and cosmological constant, respectively.

In order to deal at least partially with the fine-tuning problems stated 
above, typically quintessence models possess a special sort of scalar 
potential, `tracking attractor solutions' \cite{8}, whose cosmology is the 
same and independent of a wide set of initial conditions (around 100 orders 
of magnitude). This means that the scalar field energy density at present can 
be reached starting from a large range of initial conditions. Undoubtedly 
this fact has given an impetus for further development of quintessence models.

In spite of such a progress, 
it  turns out that the quintessence idea is difficult to implement in
the context of realistic models \cite{9,10}. The main problem besetting the
quintessential scenario has to do with the following two facts: (a) the
quintessence field should be very light, (b) time dependence 
of the expectation value of the field $\phi $, of order $M_p $ \cite{8} 
($M_p = 2.4 \times 10^{18} \; \mbox {\rm GeV}$ is the reduced Planck scale ) 
at present, is related to a phenomenon of the time variation of the 
constants of nature. In the former case, E\"{o}tv\"{o}s-type experiments put 
severe constraints on the couplings of quintessence with ordinary 
matter \cite{9}. The quintessence-exchange forces would otherwise become 
observable. A  phenomenon related to the case (a) is that the same couplings 
would unavoidably generate higher-order corrections in the quintessence 
potential, thereby spoiling the standard flatness conditions for $V(\phi )$ 
\cite{10}. 

Specifically, a coupling $\beta_{G^2} (\phi/M_p )Tr (G_{\mu \nu}G^{\mu
\nu})$, where $G_{\mu \nu}$ is the field strength tensor for QCD, is
strongly constrained \cite{9} by E\"{o}tv\"{o}s-type experiments,
\begin{equation}
\mid \beta_{G^2} \mid \leq 10^{-4} \; .
\label{form1}
\end{equation}
Similarly, a coupling such as $\beta_{F^2} (\phi/M_p )F_{\mu \nu}F^{\mu
\nu}$ will lead to the evolution of the fine-structure constant $\alpha $.
Since the present-day value for $<\phi >$ is of order $M_p $ 
\footnote{In the following, we shall always consider scalar field based
quintessence models belonging to the class for which $<\phi_0 > \sim \cal O
$ $(M_{p})$, where the subscript `0' denotes the present-day value.},
this easily
might generate corrections of order one to the gauge coupling. On the
other hand, as the time dependence of the fine-structure constant is very
strongly constrained \cite{11}, a limit \cite{9}
\begin{equation}
\mid \beta_{F^2} \mid \leq 10^{-6} \frac{M_p H_0}{<{\dot{\phi}}>} \; ,
\label{form2}
\end{equation}
can be obtained, where $H_0 $ is the present value of the expansion rate
and $<\dot {\phi}>$ is the average value over the last two
billion years. 

Note that (1) and (2) actually represent a moderate fine-tuning in a
theory. Indeed, from a traditional viewpoint, the expected values for $\beta
$s are of order unity, since they represent interactions at the Planck
scale. It should be stressed  that it is possible to alleviate the above
problem by imposing an approximate global symmetry, but this works only in
pseudo-Goldstone models of quintessence \cite{9}. A solution to the problem
in the form of the `least coupling principle' of Damour and Polyakov \cite{13} 
in string theory has recently been proposed  \cite{12}. Such a
principle was originally formulated for the string dilaton field
[or any other gauge-neutral massless scalar field (moduli) present in string
theory]. By accepting the `least coupling principle' here, we must assume
that the mechanism is also operative for quintessence since a recent analysis
\cite{14} shows that the dilaton with an exponentially decreasing potential 
cannot provide us with the negative equation of state, neither in the 
radiation-dominated nor matter-dominated era. 
Hence, the dilaton is useless for the dynamical component of
quintessence [note that Refs. \cite{12} do not aim to prove
the `least coupling principle' for
quintessence, but rather to propose a solution much in the spirit of the `least
coupling principle' of Damour and Polyakov]. Nevertheless,
observational constraints on cosmological models with quintessence arising 
from moduli fields have recently been studied \cite{15}, with the conclusion 
that for parameter 
values away from the attractor there can exist models which are consistent 
with the observational tests.     

As mentioned earlier, $<\phi_0 > \sim \cal O $ $(M_{p})$. 
This means that we need to consider the
full functions, e.g.
\begin{equation}
\beta_{F^2}(<\phi_0 >) \equiv \sum_{n=0}^{\infty} \beta_{F^2}^n
\left (\frac{<\phi_0 >}{M_p } \right )^n \; ,
\label{form3}
\end{equation}
and similarly for other $\beta $s. The `least coupling principle' states
that  string-loops effects must generate a non-trivial dependence of the
coupling functions entering the effective Lagrangian [like that in Eq.(3)],
in order to admit extrema at finite values of the $\phi $'s vev. Specifically,
the
mechanism provides that a coupling of the scalar (flavor `independence'
coupling of the string dilaton field in the scenario of Damour and Polyakov)
with the rest of the world has a common minimum close to the present value
of the $\phi $'s vev, and that the cosmological evolution naturally drives
the vev toward the minimum where the scalar decouples from matter. In
addition, even if a strictly massless dilaton exists,  
the model is able to satisfy the high-precision test of the 
equivalence principle ($\sim 10^{-12}$ level), allowing the
coupling constants that represent interactions at the Planck scale to be
species-dependent, thereby violating the equivalence principle (VEP).

In the present paper, we  study constraints from the sector of ordinary
matter and neutrinos on the quintessential Yukawa couplings. We  assume
VEP and find that such theories can be constrained substantially in both
sectors. We start by assuming that $\phi $ can couple to the
standard-model fermions via interactions of the form
\begin{equation}
\beta_{f} \frac{\phi }{M_p } {\cal L}_{Yuk}^f \;,
\label{form4}
\end{equation}
where $\beta_{f}$ is a dimensionless coupling and ${\cal L}_{Yuk}^f $ is the
gauge-invariant coupling of the Higgs doublet $\Phi $ with a
standard-model fermion. When $SU(2) \times U(1) \rightarrow U(1)_{em} $,
the field $\Phi $ acquires a non-zero expectation value, $<\Phi >$, and the
Yukawa couplings turn into 
\begin{equation}
\beta_{f} \frac{m_f }{M_p } \bar{\psi}_{f} \psi_{f} \phi \;.
\label{form5}
\end{equation}
Hence, these couplings can generate an extra Dirac mass term for the fermions.
Let us
stress that $\beta_f $ from (5) has been recently constrained for the sector
of neutrinos. Indeed, we have recently shown \cite{16} that by assuming an
interaction of quintessence with the cosmic neutrino background (CNB), it is
possible to obtain a limit on $\beta_{\nu}$
\begin{equation}
\beta_{\nu} \lsim 2 \times 10^{-3} \left ({\frac{\mbox{\rm eV}}{m_{\nu
}}} \right )^{\frac{3}{2}} \;.
\label{form6}
\end{equation}
The limit (6)  obviously depends on neutrino mass and only for large
neutrino masses, $m_{\nu } \sim 1 \; \mbox{\rm eV}$ \footnote{In $\Lambda $CDM
models, the presence of hot dark matter is no longer necessary and therefore
$\mbox{\rm eV}$ neutrinos are not needed to provide this component
\cite{17}.}, Eq. (6) represents a moderate fine-tuning in $V(\phi )$. If we set
$m_{\nu } \sim 0.05 \; \mbox{\rm eV}$ consistent with the
Super-Kamiokande experiment \cite{18}, then 
$\beta_{\nu} \lsim \cal O$ ($10^{-1}$-$1$).  
Obviously, we  need neither suppression by some
symmetry nor the `least coupling principle' to achieve these values.

Next, by considering the experimental constraints from ordinary
matter and assuming VEP, we show 
how stringent constraints on the quintessential Yukawa couplings
can be obtained. One can read off from (5) the effective
fermion mass as
\begin{equation}
m_{i}^{eff} = m_{i} + \beta_{i} \frac{<\phi_0 >}{M_p } m_{i} \;,
\label{form7}
\end{equation}
where the species-dependent parameters $\beta_{i}$ characterizes  VEP.
Since $<\phi_0 > \sim \cal O $ $(M_{p})$, the induced mass in (7) can be as
large as the bare mass if $\beta_{i}$ are of order unity. Again, we have to
consider the full functions $\beta_{i}(<\phi_0 >)$, so that the present
effective mass (7) reads
\begin{equation}
m_{i}^{eff} (<\phi_0 >) = m_{i} + \beta_{i}(<\phi_0 >) m_{i} \;.
\label{form8}
\end{equation}
If the `least coupling principle' is at work, then $\beta_{i}(<\phi_0 >)$
can stay very close to the minimum value, and hence $\beta_{i}$ can be much 
less than unity.

Now, we apply the mechanism (8) to constrain VEP from the experimental
constraint on the $K_L $-$K_S $ mass difference. By noting from (8) that the
gravitational and the mass eigenstates are identical, the effective $K_L
$-$K_S $ mass difference now reads
\begin{equation}
m_{L}^{eff}(<\phi_0 >) - m_{S}^{eff}(<\phi_0 >) = m_{L} - m_{S} +
m_{L}[\beta_{L}(<\phi_0 >) - \frac{m_{S}}{m_{L}} \beta_{S}(<\phi_0 >)] \;. 
\label{form9}
\end{equation}
Using the experimental value $(m_{L}-m_{S})/m_{L} \sim 7 \times 10^{-15}$
\cite{19},
taking an agreement of exact calculations of the mass 
difference with experiment into account, and setting $m_{S} \simeq m_{L}$ 
we get a stringent limit
\begin{equation}
\Delta \beta_{L,K} < 7 \times 10^{-15} \;,
\label{form10}
\end{equation}
where $\Delta \beta_{L,K} \equiv \beta_{L}(<\phi_0 >) - \beta_{S}(<\phi_0 >)$ 
now characterizes  VEP. It is very important to note that this bound is even
better than the most severe limit $(\sim 10^{-12} \cite{20} )$ obtained from
the extremely stringent tests of the equivalence principle. The above
bound being  independent of the absolute values of $\beta_{L}$ and $\beta_{S}$
is relevant even if we were to invoke a solution in the spirit of the
`least coupling principle' of Damour and Polyakov (with $\beta_{L,S} \ll 1
$). Finally, we would like to stress that differences of the couplings to
composite objects as in (10) are expected to contain, on the QCD basis
alone, extra small parameters in the form of the ratio of the quark masses
to the hadron mass, or the fine-structure constant. They cannot, however, be
solely responsible for the very small number in (10).

Let us now discuss a special distinction between the ``bare'' and the
``effective'' quantities entering (7) and (8), and a way to detect such a
shift as well. Since we consider here the non-universal couplings of
quintessence in the ``effective'' part of (7) and (8), they apparently
induce violation of the universality of free fall. Hence, we actually 
consider the composition-dependence of the quintessential coupling, i.e. 
its dependence on the type of matter under consideration. In such a case it 
can be easily shown \cite{13} that, in the gravitational field generated by some 
external mass, two test bodies will fall with acceleration difference 
proportional to the difference in the couplings defined as in (5). Similarly, 
the additional term in the free fall
acceleration of a body in the gravitational field of the Earth 
will be proportional to the spacetime gradient of its
effective mass. This  spacetime dependence can be clearly seen from (8).
The $\phi $-dependence of the $\beta $'s entails a corresponding $\phi
$-dependence, and therefore a spacetime dependence of $m_{i}^{eff}$. Since
different bodies $(K_L, K_S)$ have different contributions to their
$m_{i}^{eff}$, we expect the acceleration difference to differ from zero.

Giving attention to another (predominately) smoothly distributed component
in the Universe, the CNB, we now describe how sizeable fermion masses can be
generated in the matter. From a viewpoint of Thermal Field Theory (TFT)
, fermion masses within the CNB can be  generated via a
quintessence-exchange tadpole graph. As a constant independent of the
external fermion momentum, the tadpole graph at finite temperature/density
is directly related to fermion mass. It consists of a medium-induced loop
(in which real relic neutrinos are circulating) and also of a one-loop
resummed propagator for the scalar. Using the real-time version of TFT
\cite{21}, we find by explicit calculation a contribution to the induced
fermion mass as
\begin{equation}
m_{f}^{ind} \simeq 0.26 \left ( \beta_{\nu } m_{\nu } T_{\nu }^{3} \right )
\left ( \beta_f m_f  \right ) \left ( m_{\phi }^{b} M_p \right )^{-2} \;,
\label{form11}
\end{equation}
where $T_{\nu}$ and $m_{\nu }$ are the relic
neutrino temperature and the heaviest mass from the CNB, respectively.
Here $\beta_{f, \nu } \equiv \partial \; ln \; \beta_{f, \nu }(<\phi >) /
\partial (<\phi >/ M_p ) \mid_{<\phi > = <\phi_0 >}$ measures the strength
of the coupling of quintessence to the $f, \nu $-particles. 
Eq. (11) assumes that the bare  mass  for $\phi $ $(m_{\phi }^{b} \equiv
\sqrt{V''(<\phi_0 >)})$ is always larger than the quintessential thermal mass.
This is justifiably as in order for our epoch to correspond to the beginning
of slow-rolling regime, one should require that the effective mass of
fluctuations in $\phi $, $\sqrt{V''(<\phi_0 >)} \simeq H_0 $. On the other
hand, the thermal mass for $\phi $ can be $m_{\phi }^{th} \leq H_0 $.  
The properties of such graphs, but where the thermal mass dominates over the
bare mass, were discussed first time in the 
context of the dilaton-exchange gravity in \cite{22}. 
   
Next, by combining the constraint from the neutrino sector (6), with Eq.
(11), we obtain an upper limit on $\Delta \beta_{L,K}$ 
\begin{equation}
\Delta \beta_{L,K} < 1.4 \times 10^{-11} \left ( \frac{1.7 \times
10^{-4} \; \mbox {\rm eV}}{T_{\nu}} \right )^3 
\left ( \frac{m_{\nu}}{0.05 \; \mbox {\rm eV}}
\right )^{\frac{1}{2}} \;,
\label{form12}
\end{equation}
where we have used  the experimental
constraint on the $K_L $-$K_S $ mass difference.
In (12) we have put for $\beta_{\nu }$ a half of the limiting value
(6), since $m_{\phi }^{b} \simeq m_{\phi }^{th}$ near the limit.
The bound (12) becomes weaker when $\beta_{\nu }$ is decreasing. Note
that because of the negative sign of the square of  
$m_{\phi }^{th}$ for $T_{\nu}
\ll m_{\nu}$ (first ref. in \cite{12}), a more sophisticated treatment is
needed when $m_{\phi }^{b} \simeq m_{\phi }^{th}$.
Obviously, the best bound in (12) is achieved when neutrino masses are
hierarchal with the highest value of about $0.05 \; \mbox {\rm eV}$, as 
indicated by the Super-Kamiokande result. Here we stress that the effect of 
neutrino clustering, which is  not included in (12), would lead to a 
somewhat stronger
constraint. Besides, in the case when a large neutrino asymmetry exists, the
bound can be improved by about two orders of magnitude with respect to the
non-degenerate case (12). Nevertheless, the bound (10) obtained  from the
present-day value of the $\phi $'s vev, remains superior by a few orders of
magnitude.

Next, consider the sector of neutrinos. In the context of neutrino
oscillation solution to the existing neutrino anomalies, the best
constraints are expected to come from that solution with the least mass
squared difference. Hence, the obvious candidate is an explanation of the
solar neutrino data via neutrino oscillation in vacuum, with $\Delta m^2
\simeq 7.5 \times 10^{-11} \; \mbox {\rm eV}$ \cite{23}. 
Note that  VEP will be most
prominent if the oscillatory neutrinos are completely degenerate.
\footnote{The degeneracy can be protected by a presumed global inter-family
(flavor or horizontal) symmetry of leptons.} Then, by combining Eqs. (7) and
(6) we derive $\Delta m^2 $ for two degenerate-in-mass neutrinos with $m
\sim 1 \; \mbox {\rm eV}$ as 
\begin{equation}
\Delta m^2 \simeq 2  m^2 \Delta \beta_{\nu} \;,
\label{form13}
\end{equation}
where $\Delta \beta_{\nu} \equiv \beta_{\nu_2 } - \beta_{\nu_1 }$. For lighter 
neutrinos, the bound (6) is no longer restrictive and an extra term, $m^2
(\beta_{\nu_1 } + \beta_{\nu_2 }) \Delta \beta_{\nu}$, can be of the same 
order of magnitude as (13). However, in that case 
the bound on $\Delta \beta_{\nu}$ is 
weaker. One can thus conclude that the current solar neutrino data 
can probe VEP at the level
\begin{equation}
\Delta \beta_{\nu} \simeq 10^{-10} \left ( \frac{\mbox {\rm eV}}{m} \right )^2  
\; .
\label{form14}
\end{equation}
Note that the limit
(14) is much better than the limit \cite{24} obtained by comparing
neutrinos and antineutrinos from SN 1987A. The corresponding limit on
$\Delta \beta_{\nu}$ obtained from the tadpole graph is  always much weaker, 
irrespective of the neutrino mass.

Finally, we discuss constraints from the neutrinoless nuclear double
beta decay ($0 \nu \beta \beta $ decay; for review see \cite{25}). For that
purpose we need the (1,1) entry of the neutrino mass matrix in the weak
basis. For two flavors we easily obtain
\begin{equation}
M_{\nu }^{ee} = \frac{m_{\nu_1 }^{eff} + m_{\nu_2 }^{eff}}{2} + \frac{1}{2}
(m_{\nu_1 }^{eff} - m_{\nu_2 }^{eff} ) \cos{2 \theta } \;. 
\label{form15}
\end{equation}
Here, the neutrino weak interaction (flavor) eigenstates are assumed to be
linear superpositions of the mass eigenstates with a mixing angle $\theta $.
Assuming two degenerate neutrinos, the VEP part in (15) reads
$\frac{1}{2} m ~ \Delta \beta_{\nu} \cos{2 \theta }$. We see that this
contribution is very similar to that in (13). Since the  bounds from the present
$0 \nu \beta \beta $-decay experiments ($M_{\nu }^{ee} < 0.2 \; \mbox {\rm
eV}$ \cite{26}) are much less stringent than those from neutrino oscillation
experiments, one may conclude that the bound on $\Delta \beta_{\nu}$ 
from the $0 \nu \beta \beta $-decay is much weaker than those given by the
above expressions.  
   
To summarize, we have found interesting new bounds for possible
contributions of quintessence to VEP.  
We have  applied the
experimental constraints, both from the sector of ordinary matter and
neutrinos, on some quintessence-induced contributions to fermion masses. 
In some cases our bounds are
better than the upper limits from present experimental data on the
universality of the free fall. Our bounds are always better than the 
corresponding supernova limits.

{\bf Acknowledgments. } The author acknowledges the support of the Croatian
Ministry of Science and
Technology under the contract 0098011.

\end{document}